\documentclass[suppldata]{interact}

\usepackage{epstopdf}% To incorporate .eps illustrations using PDFLaTeX, etc.
\usepackage[caption=false]{subfig}% Support for small, `sub' figures and tables

\usepackage{natbib}% Citation support using natbib.sty
\bibpunct[, ]{(}{)}{;}{a}{}{,}% Citation support using natbib.sty
\renewcommand\bibfont{\fontsize{10}{12}\selectfont}% Bibliography support using natbib.sty

\theoremstyle{plain}% Theorem-like structures provided by amsthm.sty

\theoremstyle{definition}

\theoremstyle{remark}

\begin{document}

%\articletype{ARTICLE TEMPLATE}% Specify the article type or omit as appropriate

\title{Quantifying Distances Between Clusters with Elliptical or Non-Elliptical Shapes}

\author{
\name{M.~L. Wallace\textsuperscript{a}\thanks{CONTACT M.~L. Wallace. Email: lotzmj@upmc.edu} and L. Mcteague\textsuperscript{b} and J.~L. Graves\textsuperscript{c} and N. Kissel\textsuperscript{d} and C. Tortora\textsuperscript{e} and B. Wheeler\textsuperscript{f} and S. Iyengar\textsuperscript{g}}
\affil{\textsuperscript{a}Department of Psychiatry, University of Pittsburgh; \textsuperscript{b}Department of Psychiatry and Behavioral Sciences, Medical University of South Carolina; \textsuperscript{c}University of Pittsburgh Medical Center; \textsuperscript{d}Department of Statistics, Carnegie Mellon University; \textsuperscript{e}Department of Mathematics and Statistics, San Jose State University; \textsuperscript{f}School of Computing and Information, University of Pittsburgh; \textsuperscript{g}Department of Statistics, University of Pittsburgh}
}

\maketitle

\begin{abstract}
Finite mixture models that allow for a broad range of potentially non-elliptical cluster distributions is an emerging methodological field. Such methods allow for the shape of the clusters to match the natural heterogeneity of the data, rather than forcing a series of elliptical clusters. These methods are highly relevant for clustering continuous non-normal data -- a common occurrence with objective data that are now routinely captured in health research.  However, interpreting and comparing such models -- especially with regards to whether they produce meaningful clusters that are reasonably well separated -- is non-trivial. We summarize several measures that can succinctly quantify the multivariate distance between two clusters,  regardless of the cluster distribution, and suggest practical computational tools. Through a simulation study, we evaluate these measures across three scenarios that allow for clusters to differ in mean, scale, and rotation.  We then demonstrate our approaches using physiological responses to emotional imagery captured as part of the Transdiagnostic Anxiety Study, a large-scale study of anxiety disorder spectrum patients and control participants.   Finally, we synthesize findings to provide guidance on how to use distance measures in clustering applications.

\end{abstract}


\section{Introduction}
\label{sec:intro}
		Model-based clustering, also called finite mixture modeling, is an important data analytic tool because it can be used to clarify the heterogeneity of observations in a sample.  Relative to heuristic clustering approaches (e.g., k-means), model-based clustering is advantageous because it is based on an underlying statistical likelihood \citep{Aitkin} and can accommodate both elliptical and non-elliptical cluster shapes.  Recent advances in this field  include mixture models based on the generalized hyperbolic distribution (GHD)~\citep{Browne}, the skew-normal or skew-t distributions~\citep{Lee2015, Pyne2009, Sahu2003}, and the shifted asymmetric Laplace distribution~\citep{Franczak2014}, among others.   
			
Methods for interpreting clustering solutions – regardless of the underlying distribution – are also critically important.  Clustering solutions are commonly interpreted by characterizing and comparing clusters using each clustering variable. However, these univariate comparisons do not adequately quantify the \emph{multivariate} separation between each pair of clusters. A multivariate distance measure could succinctly quantify the extent to which any two clusters are distinct (i.e., with organic space between them), or whether they are highly overlapping such that they indicate a continuum. For example, a recent interesting advancement in hierarchical clustering is a statistic for measuring distances between elliptical clusters, with the purpose of determining whether they should be merged at a given stage ~\citep{Hadi}. Distances between clusters could also be informative when gauging the relative utility of two different clustering solutions, for example, one assuming elliptical clusters and one assuming non-elliptical clusters. Finally, a multivariate distance could be useful for comparing the set-up of two or more simulation studies evaluating a particular clustering methodology.    

There are several multivariate distance measures that exist within mathematical, computational, and statistical toolboxes.  However, they rarely emerge in applied research in large part because there is a paucity of practical guidance on how they might be computed and used in practice.  However, the extent to which distance metrics could be used to extract clinically or scientifically meaningful information about distances between two clusters is unknown, especially when the clusters are non-elliptical or differ in their shape.  

The primary objectives of this study are to: (1) review distance measures that may be useful for comparing clusters identified through finite mixture modeling; (2) propose practical algorithms for computing these distance measures; (3) evaluate the distance measures across a variety of simulation scenarios, including different cluster distributions and shapes; (4) demonstrate the utility of measuring distances between clusters using the Transdiagnostic Anxiety Study -- a large study of physiological reactivity to emotional imagery in a cross-diagnostic sample; and (5) synthesize findings to provide practical guidance on how to use distance measures to inform research that uses clustering.

\section{Methods}
\label{sec:meth}

\subsection{Finite Mixture Modeling}

A finite mixture model is a probabilistic approach to representing clusters within an overall sample. It can be expressed as 
\begin{equation} \label{mix}
f(\mathbf{x}| \mathbf{\Theta}) = \sum_{k=1}^{K} \pi_{k} f(\mathbf{x} | \mathbf{\Theta_{k}}), \;\; \mbox{ where } \sum_{k=1}^{K} \pi_{k} =1 \mbox{ and } \pi_{k} \geq 0,
\end{equation}
where $f(\mathbf{x} | \mathbf{\Theta_{k}})$ is the multivariate density for cluster $k$, $\pi_{k}$  are the mixing proportions, and $\mathbf{\Theta} = \{ \mathbf{\Theta_{1}}, \ldots,\mathbf{\Theta_{K}} \}$ is the set of parameters to be estimated across the clusters.  Most commonly $f(\mathbf{x} |\mathbf{\Theta_{k}})$ is taken to be a multivariate normal (MVN) density; we write $N_{d}(\boldsymbol\mu,\mathbf{C})$ for the MVN distribution in $d$ dimensions with mean $\boldsymbol\mu$ and covariance matrix $\mathbf{C}$. However, as noted previously, mixture modeling has been extended to include several other non-elliptical distributions \citep{Lee2015, Pyne2009, Sahu2003,Franczak2014}. Further recent extensions include distributions within the Multiple Coalesced Generalized Hyperbolic (MCGH)~\citep{Tortora2019} family. The distributions in this family are particularly flexible, potentially allowing each cluster distribution to have different tail behaviors. They include mixtures of MVNs as a special case. Parameters for finite mixture models can be estimated using the Expectation-Maximization (EM) algorithm or its extensions.  

\subsection{Distance Measures}

Our interest is in measuring distances between any two cluster distributions with probability density functions (pdfs) $f$ and $g$.  Useful properties of a metric for measuring distances between clusters include symmetry, scale equivariance, and computational feasibility. A metric should also \emph{ideally} be useful for both simulation studies (known cluster membership and true parameters) and applications (estimated membership and parameters), comparable across elliptical and non-elliptical cluster distributions, and/or able to suggest potential standardized effect size benchmarks. Below, we review several key distance measures, detail their computation, and describe their potential utility for measuring distances between clusters. 

\emph{Mahalanobis Distance (MD)} is one of the most commonly used metrics. MD has a closed-form expression when $f$ and $g$ are the $N_{d}(\boldsymbol\mu, \mathbf{C})$ and $N_{d}(\boldsymbol\nu, \mathbf{C})$ densities, respectively:
\begin{equation}\label{MD}
MD(f,g) = \sqrt{(\boldsymbol\mu -\boldsymbol\nu) \mathbf{C}^{-1} (\boldsymbol\mu-\boldsymbol\nu)}.
\end{equation}
MD ranges from zero to infinity, with zero indicating clusters with identical distributions and higher values indicating clusters with distributions that are farther apart. MD is considered a multivariate effect size extension of Cohen's d \citep{DelGiudice2017}.  However, a major limitation of MD for our purposes is that it is most appropriate when $f$ and $g$ are normally or elliptically distributed with equal covariance.  In situations where this assumption does not hold, some workarounds may exist; for example, using the first and second moments of a clustering distribution or computing means and pooled covariances empirically. However, when two clusters are not normally distributed and/or they do not have equal covariances, MD is not likely to have sufficient flexibility.  For example, in the case of two clusters with identical means but different covariances, computation of MD with a pooled covariance will always produce an estimate of zero. Our interest is in quantifying differences between two cluster distributions with different shapes, even if their central locations are similar. 
 
Although \emph{Bhattacharyya's Affinity (BA)} is a measure of similarity, it has many of the key properties of a useful distance measure.  For densities $f$ and $g$, BA~\citep{Bhattacharyya1943} is 
\begin{equation}\label{BAeqn1}
BA(f,g) = \int \sqrt{f(x) g(x)} dx.
\end{equation}
BA ranges from 0 to 1, with $BA=0$ when the densities $f$ and $g$ have disjoint support and $BA=1$ when they are identical.  When the densities are MVN with a common covariance matrix, BA and MD are related thus: $BA = \sqrt{-8 \ln(MD)}$.  For non-MVN pdfs $f$ and $g$, we propose that Equation (\ref{BAeqn1}) can be rewritten as

\begin{eqnarray}\label{BAeqn2} 
BA(f,g) & = & \int \sqrt{f(x) g(x)} \, dx = \int \frac{2\sqrt{f(x) g(x)}}{f(x)+g(x)} {\frac{f(x)+g(x)}{2}} \, dx \nonumber \\ & = & E\left[\frac{2\sqrt{f(U) g(U)}}{f(U)+g(U)}\right] = E \frac{1}{\cosh[\lambda(U)/2]}
\end{eqnarray}
where $U$ has the mixture pdf $(f(x) + g(x))/2$, and $\lambda(u) = \ln(f(u)/g(u))$ is the log-likelihood ratio. This novel formulation leverages the inequality $\sqrt{f(x)g(x)} \leq \frac{f(x)+g(x)}{2}$ to ensure that BA is stable and always within $[0,1]$. 

The expression in Equation (\ref{BAeqn2}) yields a simple simulation scheme to estimate $BA(f,g)$ for any defined distribution using the following equation  
\begin{eqnarray}\label{BAeqn3}
BA(f,g)&=& \frac{1}{N} \sum_{i=1}^N  2\sqrt{\frac{f(U_i)g(U_i)}{f(U_i)+g(U_i)}},
\end{eqnarray}
where $U_i=X_iT_{i}+Y_i(1-T_{i})$, $T_{i} \sim {Binom} (0.50)$, $X_i\sim f(\theta_f)$ and $Y_i\sim g(\theta_g)$. When the true probability distributions are known (e.g., simulation studies), the true parameters $\theta_f$ and $\theta_g$ can be used. In practice, estimated parameters $\hat{\theta_f}$ and $\hat{\theta_g}$ from the mixture model can be used. 

BA is symmetric and scale invariant, and  Equation (\ref{BAeqn3}) provides computational feasibility for any pdfs $f$ and $g$ from which data can be simulated. However it is not a true distance measure (it does not follow the triangle inequality) and its interpretation relates to similarity rater than distance.  Therefore, for comparability with other metrics,  we instead consider the \textit{Hellinger Distance} (HD).  HD is a true metric that is related to BA through the equation 
\begin{eqnarray}\label{HDeqn}
 HD(f,g)&  = & \sqrt{1-BA(f,g)}.
\end{eqnarray}
 HD ranges from 0 to 1, with $HD=0$ when the two distributions are identical and $HD=1$ when the two densities have disjoint support.   

The \emph{Jensen-Shannon Divergence (JS)} is a symmetrized and smoothed version of the Kullback Liebler ($KL$) divergence \citep{Kullback}, 
\begin{equation}
KL(f | g) = \int f(x) \ln \frac{f(x)}{g(x)} dx = E \ln \frac{f(X)}{g(X)}, \;\; \mbox{ where } X \sim f.
\end{equation}
JS is defined as
\begin{eqnarray}\label{JS}
JS(f,g) & = & \frac{1}{2} KL\left(f(x) \Big | \: \frac{f(x)+g(x)}{2}\right) + \frac{1}{2} KL\left(g(x) \Big | \: \frac{f(x)+g(x)}{2}\right).
\end{eqnarray}
 We call  $JSD=\sqrt{JS}$ the \emph{Jensen-Shannon Distance}; JSD is a true metric.  In practice, the JSD between two clusters with pdfs $f$ and $g$ can be computed as
\begin{eqnarray}\label{JScomp}
JSD(f,g)  =  \sqrt{
\frac{1}{2} \sum_{i=1}^N \log_2 \left(\frac{f(x_i)}{(f(x_i)+g(x_i))/2}\right) + 
\frac{1}{2}\sum_{i=1}^N \log_2 \left(\frac{g(y_i)}{(f(y_i)+g(y_i))/2}  \right) },
\end{eqnarray}
where  $X\sim f(\theta_f)$ and $Y \sim g(\theta_g)$.  
 
While $JS$ and $JSD$ are theoretically non-negative, in practice one can obtain negative JS values especially when distributions are highly overlapping and skewed (e.g., in specific cases of the MCGH distribution). A recent clever solution \citep{Nielson} addresses this limitation through the use of  an extended f-divergence that guarantees non-negative estimates.  The formulation is developed by replacing KL in Equation (\ref{JS}) by the extended KL-divergence, defined as $KL_e(f|g)=KL(f|g)+ \int (g-f) du$.  Accordingly, the \emph{Extended Jensen-Shannon Distance} ($JSD_e$) can be computed as
\begin{eqnarray}\label{JSe2}
JSD_e(f,g) = \sqrt{\frac{1}{2n}\sum_{i=1}^N \left[\frac{2f(U_i)}{s(u_i)} \log_2 \frac{2f(U_i)}{s(U_i)} +
 \frac{2g(U_i)}{s(u_i)} \log_2 \frac{2g(U_i)}{s(u_i)}\right] }
\end{eqnarray}
where $s(U_i) = f(U_i) + g(U_i)$, $U_i=X_i T_i+ Y_i(1-T_i)$, $T_i \sim {Binom} (0.50)$, $X_i \sim f(\theta_f)$ and $Y_i \sim g(\theta_g)$.  In practice, estimated parameters $\hat{\theta_f}$ and $\hat{\theta_g}$ from the mixture model can be used for both $JSD$ and $JSD_{e}$. Because $JSD$ and $JSD_e$ estimate the same quantity, and $JSD_e$ always produces non-negative estimates, we henceforth omit $JSD$ (except in Scenario 3 below to illustrate the shortcoming of $JSD$).

The \emph{Wasserstein Distance}, also known as Mallows Distance \citep{Mallows, Rachev} and related to the Earth Mover's Distance \citep{Rubner1,Rubner2}, can be conceptualized as the cost of transporting a pile of sand ($f$) to cover a sinkhole ($g$). It can be expressed as 
\begin{eqnarray}\label{WD}
WD(f,g)=& \min_{H} \left\{ \left( E \left\|X-Y\right\|^p \right)^{1/p} : (X,Y) \sim H, X \sim f, Y \sim g\right\}
\end{eqnarray}
where $X\sim f(\theta_f)$, $Y\sim g(\theta_g)$, $(X,Y)\sim H(\theta_H)$, $(X,Y \in \mathbb{R}^d)$, and $p$ is such that the moments are finite (usually $p=1$ or 2). Equation (\ref{WD}) can be interpreted as the $p^{th}$ root of the total cost of optimally transporting $f$ to $g$, where the cost of transporting a unit of mass from $X$ to $Y$ is the $p^{th}$ power of the Euclidean distance $\|X-Y\|^{p}$.  The \texttt{R} package \texttt{transport}\citep{WDinR} facilitates the computation of WD in practice, as detailed in Section \ref{sec:sim}. 

The \emph{Average Between} (AB) and \emph{Separation Index} (SI) are two empirical indices that can be computed using the \texttt{cluster.stats()} function within the \texttt{fpc} package~\citep{fpc} in R.  AB is computed by taking the average of distances between points in each cluster.  SI is computed by first measuring distances for every point to the closest point not in the same cluster. The mean of the smallest proportion (e.g., 10\%) is then computed \citep{Hennig2014}. Thus, AB and SI differ fundamentally from other distance measures we consider because they summarize individual distances between points and do not require parameters to be known (or estimated). Further computational details are provided in Section \ref{sec:sim}.  

\section{Simulation}
\label{sec:sim}

%\addtolength{\textheight}{.5in}%

We evaluated the metrics using three simulation scenarios.   In Scenario 1, we considered two bivariate-normal  clusters with equal covariance but different means. In Scenario 2, we considered two bivariate-normal clusters with identical means but different scales.  In Scenario 3, we considered two bivariate clusters  generated from multiple generalized hyperbolic (MGH) distributions. The means and scales were identical in both clusters; however the magnitude and direction of skewness varied.  These three scenarios were motivated by the application (Section \ref{sec:app}), which exemplifies how clusters can vary in their mean levels as well as their rotation, skewness, and scale. Parameter values were selected to study the behaviors of the metrics across their full ranges (where possible).  All simulations were performed using R version 4.0.2 \citep{R}. 

\emph{True Distances.} For Scenarios 1 and 2 (MVN distributions) we computed true distances based on the known parameters ($HD, JSD_e, WD, MD$) or known cluster membership ($AB$ and $SI$).  For Scenario 3 (MGH distribution) we evaluated the same metrics, with the exception of $MD$ because it is based on the MVN distribution.  $JSD$ was also extracted for Scenario 3 for the purpose of comparing $JSD$ and $JSD_e$ in the context of non-elliptical clusters. 

\emph{Empirical Distances.} To reflect a more realistic scenario where the true cluster distributions and assignments are unknown, we simulated $100$ datasets ($N=1000$, equal cluster sizes) for each parameter combination within each scenario.  For Scenarios 1 and 2 we fit MVN mixture models using the \texttt{Mclust} package~\citep{mclust} and \texttt{mclust} function. For Scenario 3 we fit MGH mixture models using using \texttt{MixGHD}~\citep{MixGHD} package and the \texttt{MGHD} function.  When fitting the mixture models, we considered only two clusters. We then extracted empirical parameter estimates from the mixture models to compute $HD$, $JSD_e$, $WD$, and $MD$, and extracted cluster assignments to compute $AB$ and $SI$. 

\emph{Computing Details.}  $MD$ was computed with the true or estimated parameters using Equation (\ref{MD}) and a pooled covariance.  $HD$ was computed using Equations (\ref{BAeqn3}) and (\ref{HDeqn}) with $N=1000$. $JSD_e$ was computed using Equation (\ref{JSe2}) with $N=1000$. For $WD$ we drew $N=1000$ samples from each bivariate cluster distribution, denoted $X$ and $Y$ respectively.  Within the \texttt{transport} package we used the \texttt{pp} function to create discretized measures of $X$ and $Y$ with fixed mass at finitely many locations, denoted $X_d$ and $Y_d$ respectively. We then computed WD using the \texttt{wasserstein} function to develop the ``transport map'' between $X_d$ and $Y_d$, with  the number of finite moments set as $p=2$.  For true $AB$ and $SI$, we simulated $100$ datasets ($N=1000$, equal cluster size) using the true parameters, scaled the data, computed the index based on the true cluster membership, and averaged across the simulated data sets. For the estimated $AB$ and $SI$, we used the observed cluster membership from the mixture model and the \texttt{cluster.stats()} function in the \texttt{fpc} package. 

\emph{Cluster Recovery.}  We  computed the Adjusted Rand Index (ARI)\citep{Steinley} to evaluate cluster recovery for the mixture models. Based on recommendations~\citep{Steinley} we used ARI cutoffs to indicate poor  $(-1,0.65)$, moderate $[0.65-0.80)$, good $[0.80-0.90)$ and excellent $[0.90,1)$ recovery. 

\subsection{Scenario 1: Bivariate Normal Clusters with Varying Means}

\begin{figure}
\centering
\resizebox*{15cm}{!}{\includegraphics{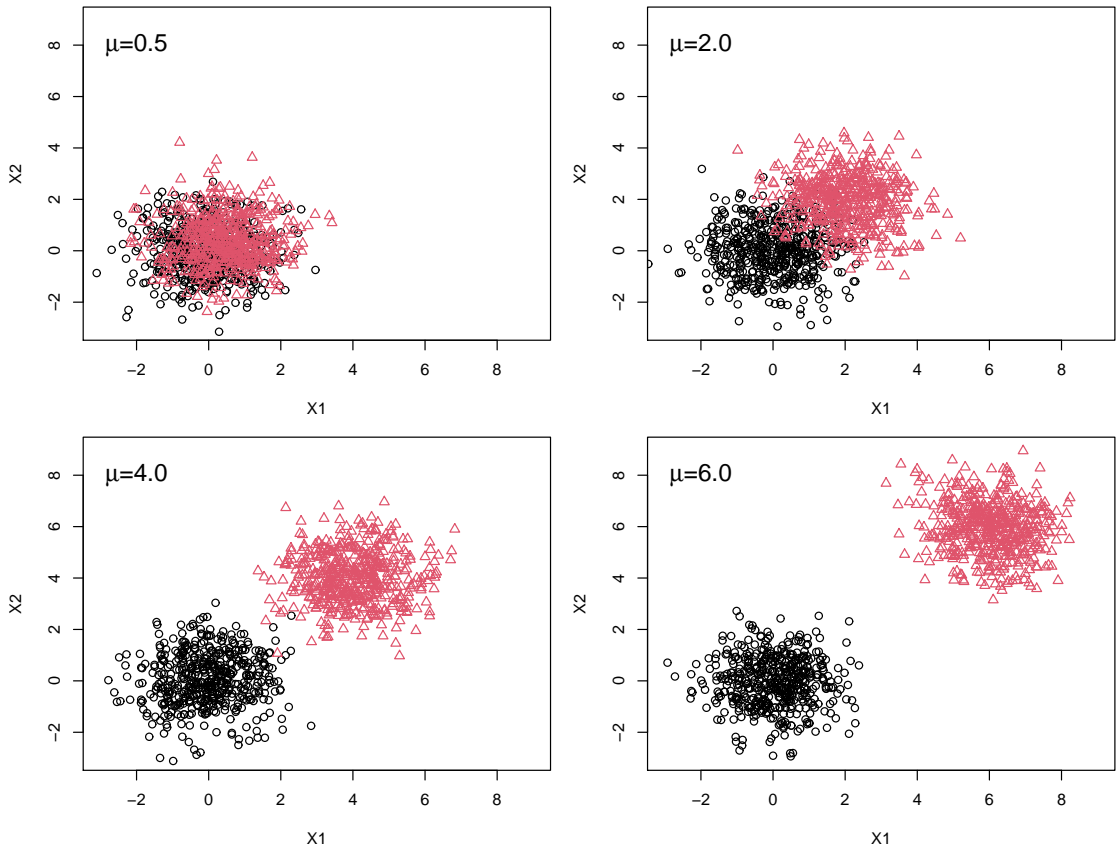}}\hspace{5pt}
\caption{Example Data (Scenario 1).} \label{ExampleSim1}
\end{figure}

We simulated two bivariate normal distributions $f \sim N_{2}([0,0]',\mathbf{C})$ and $g \sim N_{2}([\mu,\mu]',\mathbf{C})$, allowed  $\mu$ to range from 0 to 6 in steps of 0.50  (selected to display the full range of the distance measures), and set $\mathbf{C} = 0.30 \mathbf{I}$ where $\mathbf{I}$ is the identity.  See Figure \ref{ExampleSim1} for an example of the simulated data.  

%\begin{figure}[H]
%\center
%\caption{Example Data (Scenario 1).}
%\epsfig{file=Example_Simulation1.eps,width=8cm,height=8cm}\label{ExampleSim1}
%\end{figure}

\begin{figure}
\centering
\resizebox*{15cm}{!}{\includegraphics{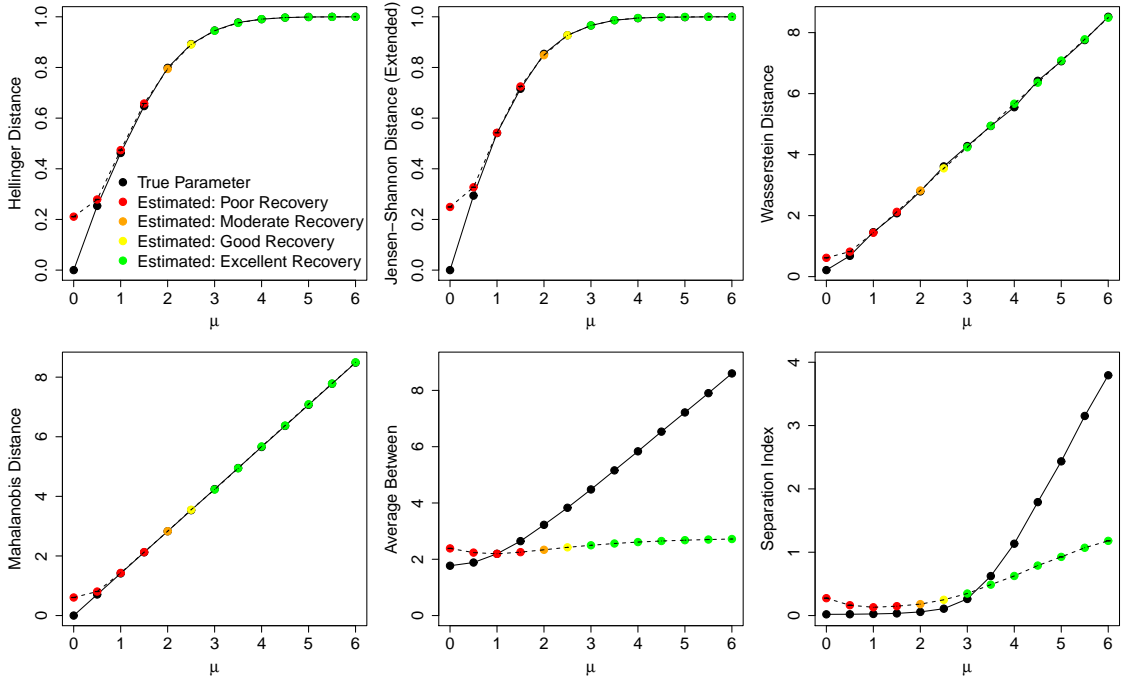}}\hspace{5pt}
\caption{True and Empirical Distances (Scenario 1).} \label{Sim1}
\end{figure}

\emph{True Distances}. Results are provided in Figure \ref{Sim1} (solid black lines).  $MD$ and $WD$ are nearly identical and increase linearly as the two clusters become farther apart. $AB$ also increases linearly, however the true AB is not zero even for identical distributions.  $HD$ and $JSD_e$ are also nearly identical; they are highly sensitive to small changes when the clusters are overlapping but have limited sensitivity when clusters are further apart. In contrast, $SI$ has limited sensitivity when cluster are overlapping but is highly sensitive when clusters are further apart.  

%\begin{figure}[H]
%\center
%\caption{True and Empirical Distances (Scenario 1).}
%\epsfig{file=Est_true_Scenario1.eps,width=15cm,height=8cm}\label{Sim1}
%\end{figure}

\emph{Empirical Distances}.  Empirical distances are displayed by the dashed lines in Figure \ref{Sim1}, with the colored points coded based on the ARI for that scenario. For all scenarios except the most highly overlapping ($\mu \leq 0.5$, poor cluster recovery), the estimated $HD$, $MD$, $WD$, and $JSD_e$ are accurate.  In contrast, the estimated $AB$ and $SI$ performed poorly; they significantly underestimated the distance when cluster recovery was excellent. 

\subsection{Scenario 2: Bivariate Normal Clusters with Varying Scales}

\begin{figure}
\centering
\resizebox*{15cm}{!}{\includegraphics{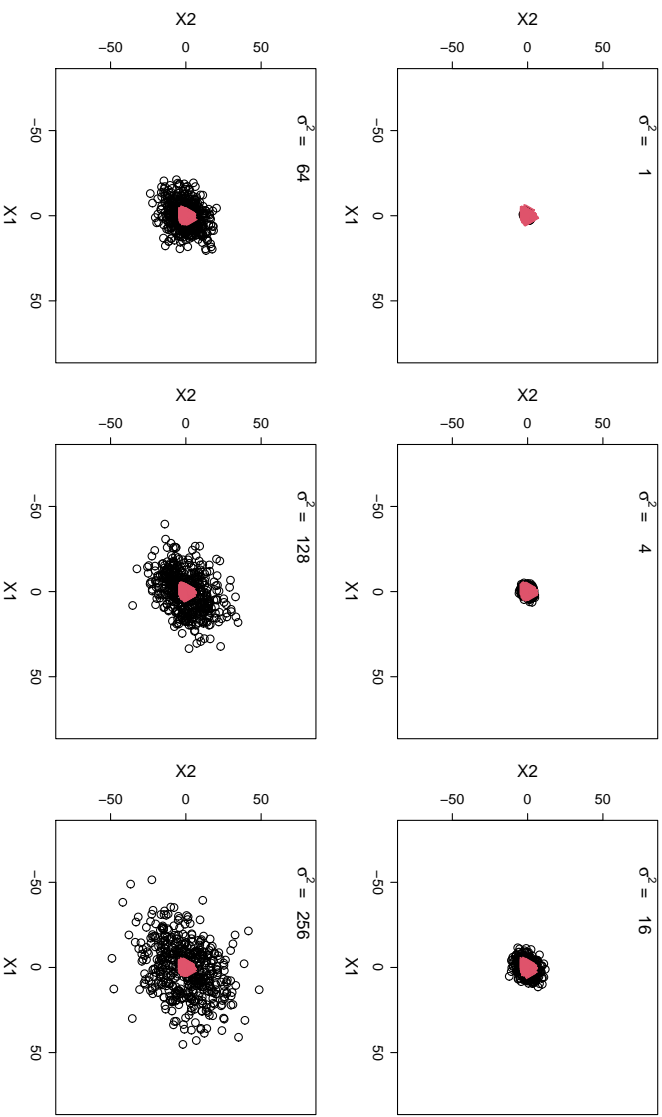}}\hspace{5pt}
\caption{Example Data (Scenario 2).} \label{ExampleSim2}
\end{figure}

We simulated two bivariate normal distributions $f \sim N_{2}([0,0]',\mathbf{C_1})$ and $g \sim N_{2}([0,0]',\mathbf{C_2})$ where $\mathbf{C_1} = 0.30 \mathbf{I}$ and $\mathbf{C_2}=0.30 \sigma^2 \mathbf{I}$ are proportional to the identity matrix $\mathbf{I}$, with $\sigma^2$ doubling among $\{1, 2, 4, \ldots, 512 \}$.  See Figure \ref{ExampleSim2} for an example of the simulated data.  

%\begin{figure}[H]
%\center
%\caption{Example Data (Scenario 2).}
%\epsfig{file=Example_Simulation3.eps,width=12cm,height=9cm}\label{ExampleSim2}
%file was named when simulation 2 was simulation 3 and then I reversed them%
%\end{figure}

\begin{figure}
\centering
\resizebox*{15cm}{!}{\includegraphics{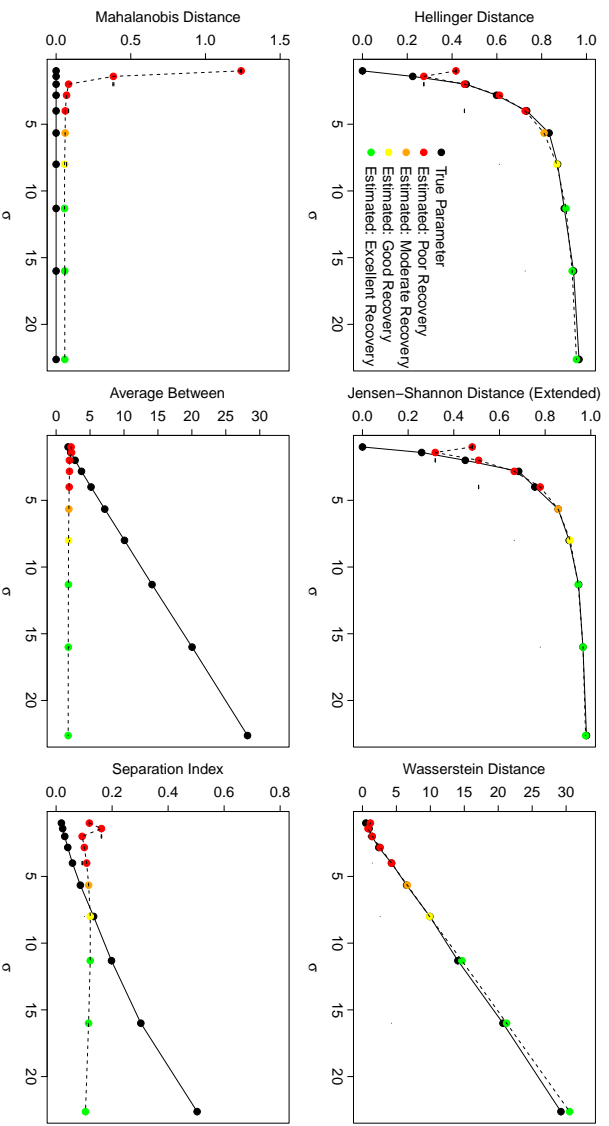}}\hspace{5pt}
\caption{True and Empirical Distances (Scenario 2).} \label{Sim2}
\end{figure}

\emph{True Distances.}  Results are provided in Figure \ref{Sim2} (solid black lines).  $WD$, $AB$, and $SI$ increase relatively linearly with larger differences in scale. $HD$ and $JSD_e$ are more sensitive to small scale differences but less sensitive to large scale differences. $MD$ is fixed at $0$, which is expected because the means are identical.  

%\begin{figure}[H]
%\center
%\caption{True and Empirical Distances (Scenario 2).}
%\epsfig{file=Scenario3_All.eps,width=15cm,height=10cm}
%file was named when simulation 2 was simulation 3 and then I reversed them%
%\label{Sim2}
%\end{figure}

\emph{Empirical Distances.}   Empirical distances are displayed by the dashed lines in Figure \ref{Sim2}, with the colored points coded based on the ARI. Estimated $WD$ is accurate across all scenarios, even with poor cluster recovery. Estimated $HD$ and $JSD_e$ are accurate except for in the two most highly overlapping scenarios ($\sigma^2={1, 2}$, poor cluster recovery).  Estimated $AB$, $SI$, and $MD$ cannot cannot distinguish any differences across the scenarios, even when cluster recovery is excellent. 

\subsection{Scenario 3: MGH Distribution with Varying Rotation and Skewness}

\begin{figure}
\centering
\resizebox*{15cm}{!}{\includegraphics{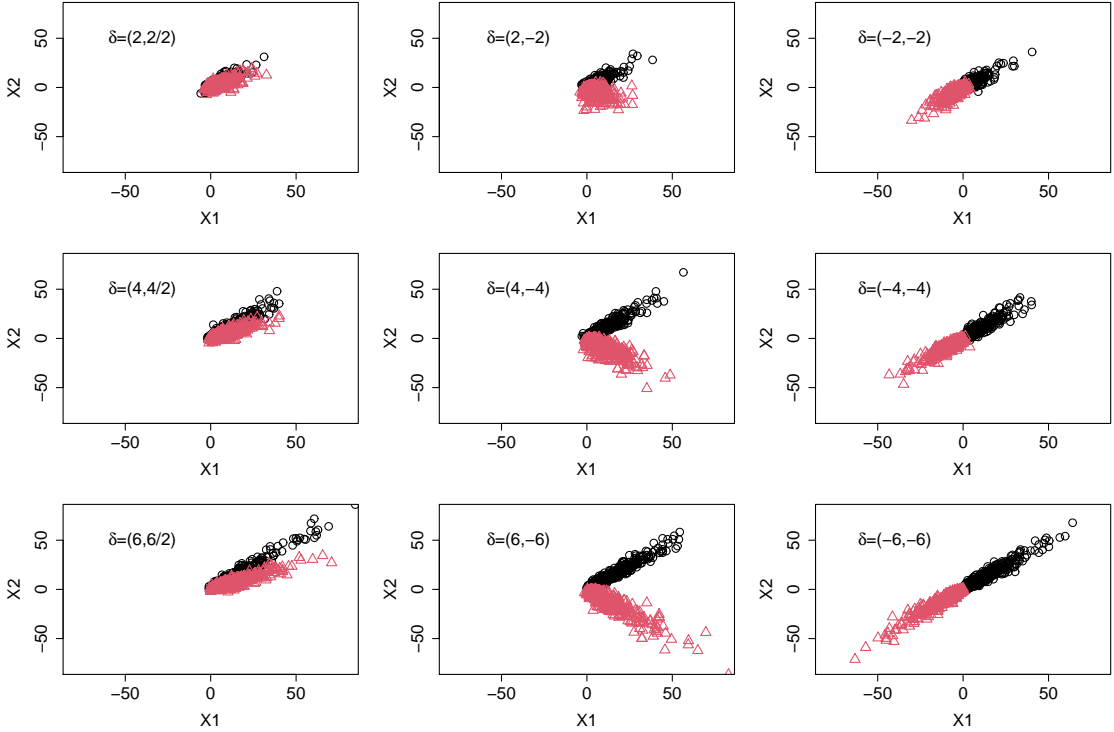}}\hspace{5pt}
\caption{Example Data (Scenario 3).} \label{ExampleSim3}
\end{figure}

We simulated two clusters, each following a bivariate generalized hyperbolic distribution. The locations and scales of the two clusters were identical, such that $\mathbf{\mu_{1}} = \mathbf{\mu_{2}} = \mathbf{0}$ and
\[ \mathbf{\Sigma_{1}} = \mathbf{\Sigma_{2}} = \left[\begin{array}{cc}\nu & 0.3 \nu \\ 0.3\nu & \nu \end{array}\right] \]
with $\nu=4$. Concentration parameters $\theta$ and $\lambda$ were  set to 1 for both clusters.  For cluster 1 we set the magnitude and direction of skewness as $\mathbf{\delta_{1}} = (d,d)$, with $d= {2,4, 6 }$.  For cluster 2 we varied the skewness every $22.5^{\circ}$ around the circle coordinates with radius $d$ until the clusters were in opposite directions (180°) at $(-d, -d)$.  Example data are displayed in Figure \ref{ExampleSim3}.  

%\begin{figure}[H]
%\center
%\caption{Example Data (Scenario 3).}
%\epsfig{file=Example_Simulation2.eps,width=10cm,height=10cm}\label{ExampleSim3}
%\end{figure}

\begin{figure}
\centering
\resizebox*{15cm}{!}{\includegraphics{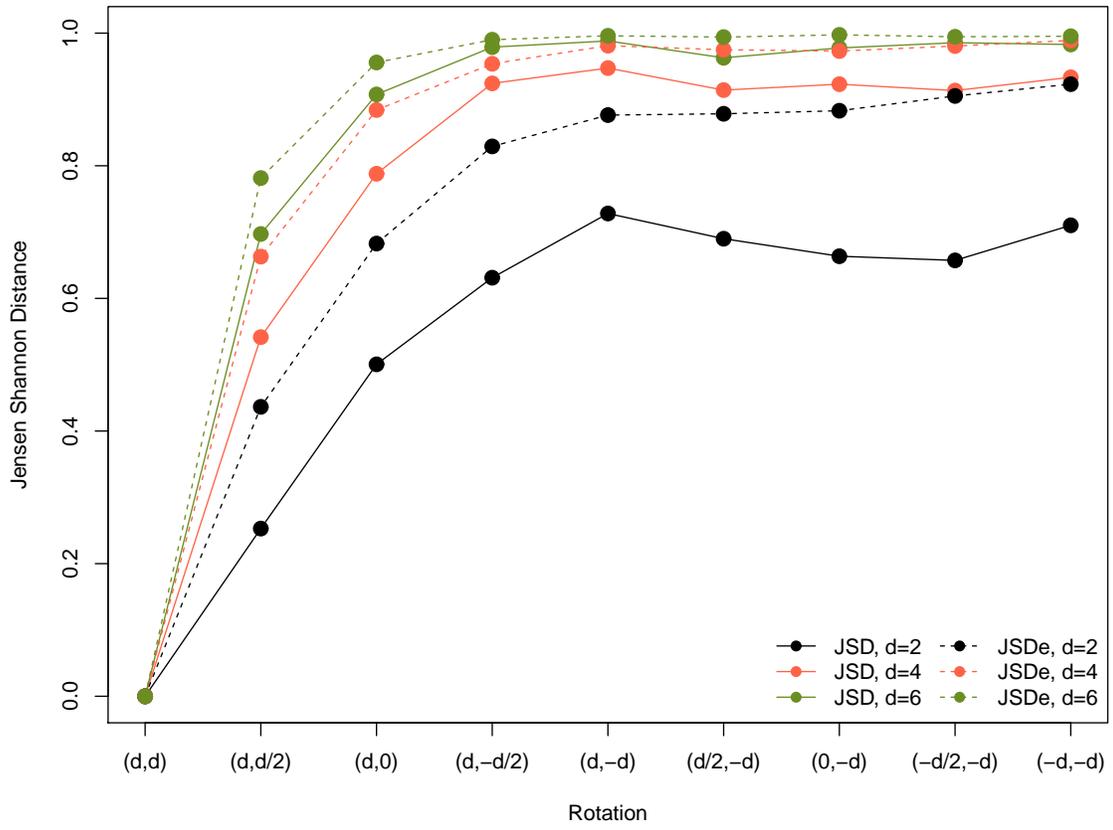}}\hspace{5pt}
\caption{Standard vs. Extended Jensen Shannon Distance (Scenario 3).} \label{figJSD3}
\end{figure}

\emph{True Distances.} We first examined $JSD$ and $JSD_e$ across the various parameter combinations (Figure \ref{figJSD3}). $JSD$ and $JSD_e$ do not correspond well, especially in situations where the clusters are highly overlapping but not identical (e.g. $d=2$).  $JS$ (used to compute $JSD$) can have components whose sum can be negative. In the case of the MGH distribution, which has large observations with appreciable probability, these negative components can be extreme and pull the value lower.  In contrast, $JSD_e$ by definition cannot have negative components within the square root, which ultimately results in a larger $JSD_e$ relative to $JSD$.  We only further consider $JSD_e$ within this Scenario. 

%\begin{figure}[H]
%\center
%\caption{Standard vs. Extended Jensen Shannon Distance (Scenario 3)}
%\epsfig{file=JS_Comparison_Scenario2.eps,width=12cm,height=8cm}\label{figJSD3}
%\end{figure}

\begin{figure}
\centering
\resizebox*{15cm}{!}{\includegraphics{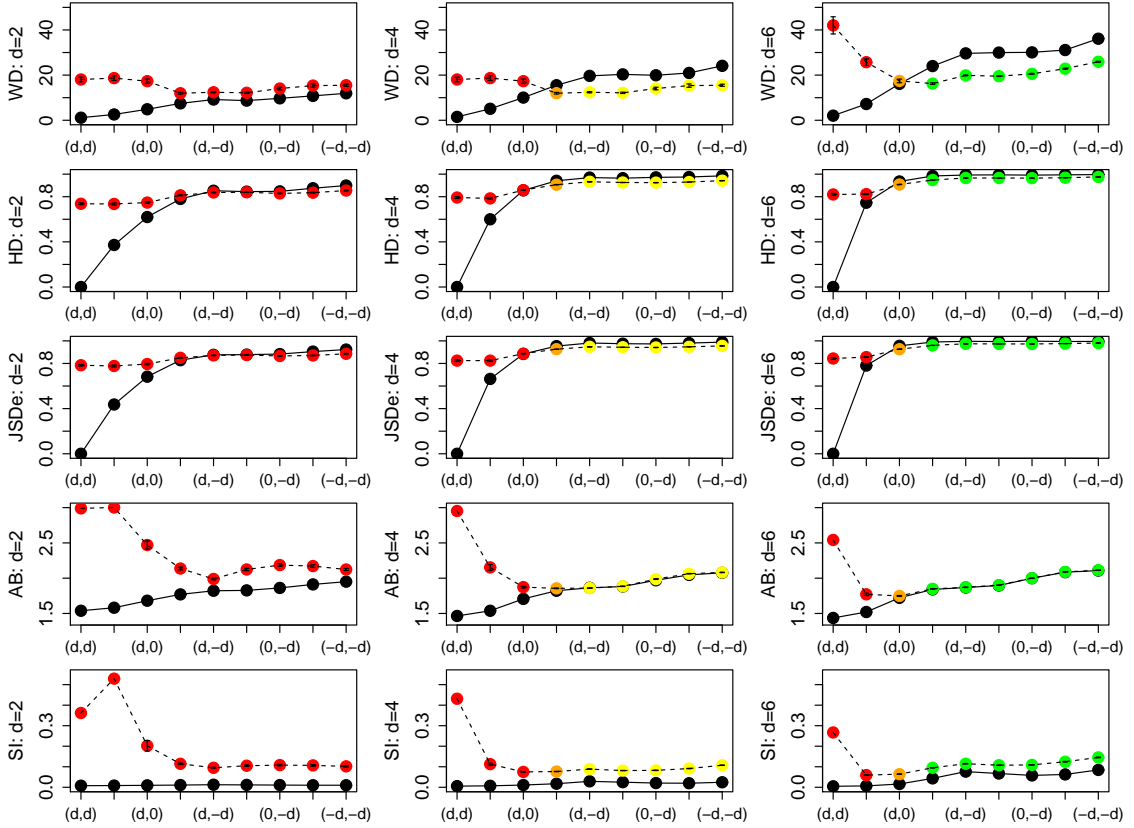}}\hspace{5pt}
\caption{True (solid lines) and empirical (dashed lines) distances for Scenario 3. Color corresponds to cluster recovery (ARI): poor is red, orange is moderate, yellow is good, green is excellent. } \label{fig3}
\end{figure}

The solid black lines in Figure \ref{fig3} display the true distances across parameter combinations.  Both $WD$ and $AB$ increase linearly as the second cluster moves clockwise around the circle to $(d,-d)$ (a $90^{\circ}$ rotation from cluster 1). $WD$ has two additional beneficial properties not observed with $AB$: (1) it increases with a higher magnitude of skewness $d$ and (2) it is $0$ when the two clusters are identical. $HD$ and $JSD_e$ again perform similarly to one another. They are both sensitive to the smaller initial rotations but have limited sensitivity to changes in rotation past $(d, -d)$ ($90^{\circ}$ rotation). At $(d, d)$, where the two clusters are identical, both $JSD_e$ and $HD$ are appropriately $0$;  they also both asymptotically approach $1$ as the rotation for cluster 2 nears $(-d,-d)$ (a $180^{\circ}$ rotation from cluster 1). $SI$ displays limited change except for with the largest magnitude of skewness ($d=6$). 

%\begin{figure}[H]
%\center
%\caption{True (solid lines) and empirical (dashed lines) distances for Scenario 3. Color corresponds to cluster recovery (ARI): poor is %red, orange is moderate, yellow is good, green is excellent. }%
%\epsfig{file=Scenario2_All.eps,width=15cm,height=12cm}
%\label{fig3}
%\end{figure}

\emph{Empirical Distances.} Empirical distances are displayed by the dashed lines in Figure \ref{fig3}, with the colored points coded based on the ARI.  $WD$ over-estimates the distance when recovery is poor, but under-estimates the distance when recovery is good-to-moderate.   $HD$ and $JSD_e$ over-estimate the distance with $\leq 22.5^{\circ}$ rotation difference between clusters.  With at least a $45^{\circ}$ rotation difference, $HD$ and $JSD_e$ are relatively accurate; despite their general lack of sensitivity to changes in rotation within this range. $AB$ drastically over-estimates smaller rotation differences, especially when $d=2$.  However, $AB$ was surprisingly accurate with at least moderate cluster recovery and also increased somewhat linearly with added rotations.  Finally, $SI$ consistently over-estimated distances, especially in the scenarios corresponding to poor cluster recovery. 

\section{Application}
\label{sec:app}

Data from the Transdiagnostic Anxiety Spectrum study were collected by Peter Lang, Ph.D. and colleagues at the University of Florida Center for the Study of Emotion and Attention (MH082702; MH72850). The analytic sample employed here includes $N=505$ adults across eight different anxiety disorders plus major depressive disorder and healthy controls.  Participants in this study listened to  narratives varying in emotional imagery while five physiological responses were measured: Corrugator response (COR),  orbicularis electromyography (ORB),  startle reflex (STL), skin conductance response (SCR), and heart rate (HR). In this illustration we focus on identifying cross-diagnostic subtypes with similar physiological responses to pleasant imagery relative to neutral imagery. Measures were computed as the pleasant physiological response minus the neutral physiological response.   

\subsection{Selecting the Number of Clusters}

Using \texttt{R} we fit both MVN (\texttt{Mclust} package) and MCGHD (\texttt{MixGHD} package) mixture models to the five physiological responses to pleasant versus neutral imagery, considering 1-5 clusters.  For MVN we also considered the large array of covariance types that are available through the package.  To select the optimal model for \texttt{Mclust} we selected the one that maximized the $-BIC$, which is conventional for this approach.  To select the optimal model for \texttt{MixGHD}, we used a more nuanced approach because the models are more complex and have a much larger number of parameters to estimate. Therefore, we considered the Bayesian Information Criteria (BIC), Akaike Information Criterion ($AIC$), and Integrated Complete-Data Likelihood ($ICL$). We prioritized clustering solutions that maximized the $-AIC$ and $-BIC$ and/or which preceded a drop in ICL.  We identified 3 MCGHD clusters using \texttt{MixGHD} $(C1 - C3)$ and 4 MVN clusters using \texttt{Mclust} $(C1 - C4)$. 

\subsection{Cluster Characterization using Descriptive Statistics}

We used descriptive statistics (median [Q1, Q3]) to summarize the physiological responses in each cluster (Table \ref{tab:mclust}). We also used Kruskal-Wallis tests to determine whether the responses differed by cluster. 

Based on the MVN mixture model solution, participants in $C4$ have the strongest response to pleasant (vs. neutral) imagery on COR, ORB, STL, and HR, while participants in $C1$ have the strongest response on SCR.  Participants in clusters $C2$ and $C3$ have lower-moderate responses. SCR, HR, and ORB differed significantly across the four clusters identified using the MVN mixture model.  

Based on the MCGHD mixture model solution, participants in $C1$ have the strongest response on STL, SCR, and HR, participants in $C2$ have the strongest response on COR and ORB, and participants in $C3$ have lower responses.  All five measures differed significantly across the three clusters identified using the MCGHD  mixture model. 

\begin{sidewaystable}
 \tbl{Median (Q1, Q3) of physiological characteristics by cluster.}
  {\begin{tabular}{|l|c|c|c|c|c|}
\hline
	&	Startle Response	&	Skin Conductance Response	&	Heart Rate	&	Corrugator Response	&	Orbicularis Response	\\ \hline
Total  (N=505)	&	1.53 (-1.29, 5.27)	&	0.00 (-0.00, 0.02)	&	0.39 (-0.72, 1.52)	&	-0.05 (-0.37, 0.23)	&	0.52 (0.06, 1.19)	\\ \hline
\multicolumn{6}{|l|}{\textit{MCGHD Mixture Model}} \\ \hline	
C1  (N=96)	&	4.00 (0.59, 8.20)	&	0.02 (0.00, 0.05)	&	1.92 (1.44, 2.99)	&	0.05 (-0.07, 0.27)	&	0.68 (0.31, 1.28)	\\ \hline
C2  (N=132)	&	0.97 (-0.71, 3.63)	&	0.00 (0.00, 0.01)	&	0.35 (-0.52, 1.15)	&	0.26 (-0.86, 1.06)	&	1.17 (0.39, 2.53)	\\ \hline
C3  (N=277)	&	1.17 (-2.58, 4.96)	&	0.00 (-0.01, 0.01)	&	-0.16 (-1.15, 1.04)	&	-0.13 (-0.41, 0.04)	&	0.25 (-0.02, 0.76)	\\ \hline
\multicolumn{6}{|l|}{\textit{MVN Mixture Model}} \\ \hline	

C1  (N=226)	&	1.55 (-2.67, 6.20)	&	0.02 (-0.02, 0.04)	&	0.53 (-0.74, 1.74)	&	-0.06 (-0.33, 0.14)	&	0.52 (0.08, 1.17)	\\ \hline
C2  (N=90)	&	0.59 (-1.32, 3.25)	&	0.00 (0.00, 0.01)	&	0.13 (-1.14, 1.08)	&	0.08 (-1.04, 0.98)	&	0.97 (0.03, 2.13)	\\ \hline
C3  (N=137)	&	1.81 (-0.43, 4.78)	&	0.00 (0.00, 0.01)	&	0.24 (-0.63, 1.26)	&	-0.06 (-0.23, 0.08)	&	0.27 (-0.02, 0.55)	\\ \hline
C4  (N=52)	&	2.58 (-0.84, 10.88)	&	0.00 (-0.02, 0.04)	&	1.38 (-0.48, 2.63)	&	0.14 (-1.05, 1.53)	&	2.09 (0.59, 3.90)	\\ \hline
\end{tabular}}\label{tab:mclust}
\end{sidewaystable}

\subsection{Cluster Contour Plots}

\begin{figure}
\centering
\resizebox*{15cm}{!}{\includegraphics{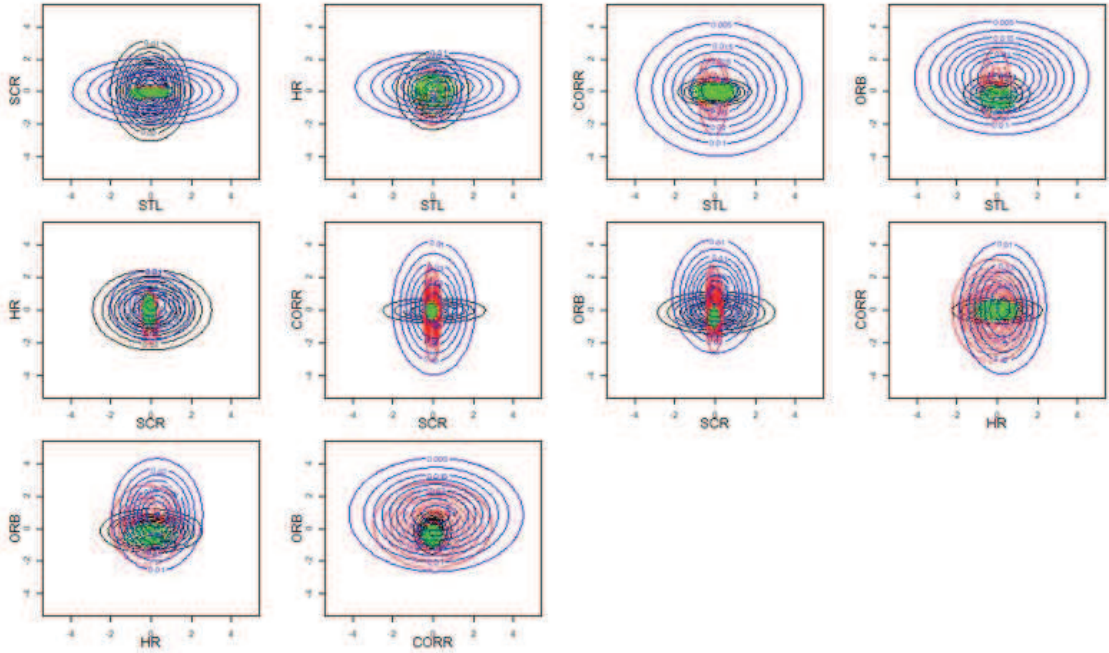}}\hspace{5pt}
\caption{Contour plots for MVN clusters. Cluster 1 = black; Cluster 2 =  red; Cluster 3 = green; Cluster 4 = blue.} \label{MVN_contour}
\end{figure}

Figure \ref{MVN_contour} illustrates elliptical clusters generated using parameter estimates from the MVN mixture model. $C4$ has a large scale and encompasses the other clusters, despite the relatively small sample size ($N=52$).  In contrast, $C3$ has a very small scale relative to the other clusters. Perhaps as a result of these notable differences in scale, it is difficult to graphically distinguish mean differences between clusters. 

%\begin{figure}[H]
%\center
%\caption{Contour plots for MVN clusters. Cluster 1 = black; Cluster 2 =  red; Cluster 3 = green; Cluster 4 = blue.}
%\epsfig{file=MVN_Application_Contour.eps,width=15cm,height=10cm}
%\label{MVN_contour}
%\end{figure}

\begin{figure}
\centering
\resizebox*{15cm}{!}{\includegraphics{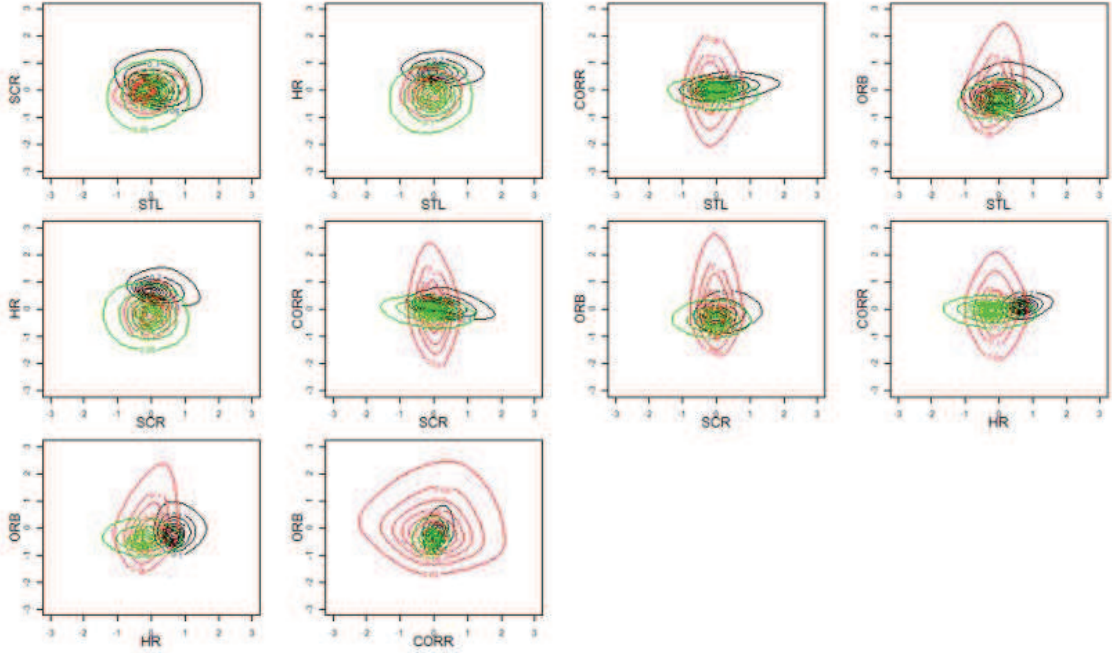}}\hspace{5pt}
\caption{Contour plots for MCGHD clusters. Cluster 1 = black; Cluster 2 =  red; Cluster 3 = green.} \label{MGHD_contour}
\end{figure}

Figure \ref{MGHD_contour} illustrates the clusters generated using the parameter estimates from the MCGHD mixture model.  Clusters are non-elliptical with differences in location, scale, and rotation. For example, in plots of COR vs. HR and ORB vs. HR, $C2$ is rotated $90^{\circ}$ from $C3$ and $C1$, and $C1$ generally has higher HR levels than $C2$ and $C3$. $C2$ also has a larger scale on COR and ORB relative to the other clusters.  

%\begin{figure}[H]
%\center
%\caption{Contour plots for MCGHD clusters. Cluster 1 = black; Cluster 2 =  red; Cluster 3 = green.}
%\epsfig{file=MCGHD_Application_Contour.eps,width=15cm,height=10cm}
%\epsfig{file=HD_Scenario2.eps,width=15cm,height=8cm}
%\label{MGHD_contour}
%\end{figure}

\subsection{Distance Measures}

Although  summary statistics and plots  important tools for cluster interpretation, they cannot succinctly summarize: (1) which clusters are most separate from other clusters within a given solution; or (2) whether the MVN or MCGHD solution suggests pairs of clusters that are more distinct. The proposed distance measures can fill these gaps.  For each pair of clusters and each clustering approach (MCGHD, MVN), we computed four metrics previously considered in the simulation study ($HD$, $JSD_e$, $WD$, $AB$, and $SI$). However, we gave $HD$, $JSD_e$, and $WD$ more consideration because of the relatively poor performance of $AB$ and $SI$ in the simulation study. $MD$ was not considered because its computation is not meaningful for the MCGHD model and it cannot distinguish differences in scale.   

%\begin{figure}[H]
%\center
%\caption{Distances between pairs of clusters. The top row reflects pairwise distances from the MVN model. The bottom row reflects pairwise distances from the MCGHD model}
%\epsfig{file=pln_mclust_ghd_distances_radial.eps,width=17cm,height=7cm}
%\label{heatmap1}
%\end{figure}

%\begin{figure}
%\centering
%\resizebox*{20cm}{!}{\includegraphics{pln_mclust_ghd_distances_radial.eps}}\hspace{5pt}
%\caption{Distances between pairs of clusters. The top row reflects pairwise distances from the MVN model. The bottom row reflects pairwise distances from the MCGHD model} \label{heatmap1}
%\end{figure}

%\setcounter{figure}{1}
\begin{sidewaysfigure}
\centerline{
\epsfxsize=9.5in
\epsfbox{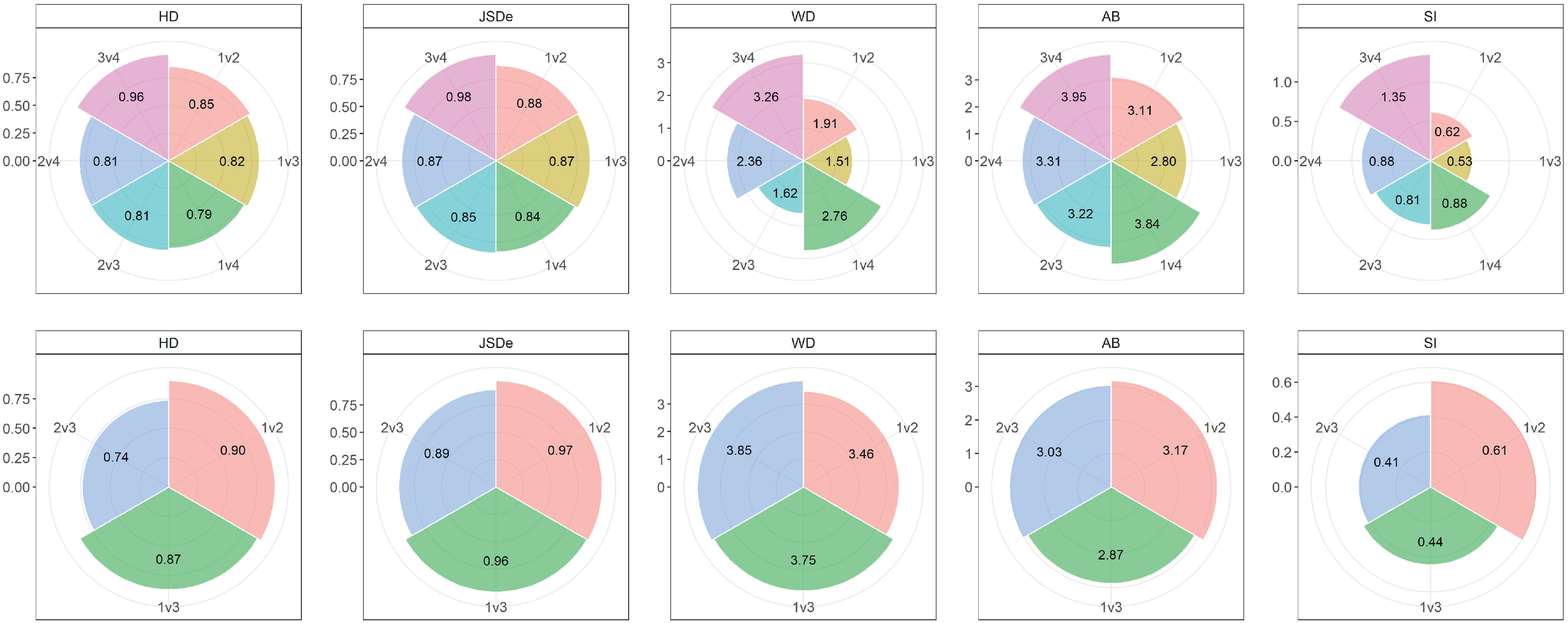}
}
\caption{Distances between pairs of clusters. The top row reflects pairwise distances from the MVN model. The bottom row reflects pairwise distances from the MCGHD model.}
\label{heatmap1}
\end{sidewaysfigure}

The top row of plots in Figure \ref{heatmap1} display pairwise cluster distances from the MVN model. All five metrics indicate that $C3$ and $C4$ have the greatest distance between them, and that $C2$ is not particularly distinct from other clusters. $SD$, $AB$, and $SI$ indicate that $C1$ and $C4$ are well-separated. 

The bottom row of plots in Figure \ref{heatmap1} displays pairwise cluster distances from the MCGHD model. The distances between each pair of clusters are relatively large when quantified by $HD$, $JSD_e$, $WD$, and $AB$. There is some inconsistency regarding which specific pair has the largest distance, with $HD$, $JSD_e$, $AB$, and $SI$ indicating $C1$ and $C2$ have the largest distance, and $WD$ indicating that $C2$ and $C3$ have the largest distance. Despite this incongruity, $C2$ still emerges as the cluster that is most distinct from the other clusters.  

When comparing each measure across MCGHD and MVN clustering solutions, distances between MCGHD clusters are overall larger; however some some specific pairwise distances in the MVN solution are also large (e.g., between $C3$ and $C4$).  It is also noteworthy that $WD$ provides a high level of sensitivity to distinguishing cluster differences among pairs, whereas $HD$ and $JSD_e$ have limited ability to distinguish such differences.  This is consistent with the simulation study results, which showed $HD$ and $JSD_e$ are less sensitive to detecting differences between larger distances.

\section{Discussion}
\label{sec:disc}

Interpretation of a clustering solution is often the most challenging aspect of a cluster analysis. Metrics that aid in determining whether clusters are meaningfully distinct from one another can facilitate this process.  Given the expanding field of finite mixture modeling with non-elliptical clusters, it is essential to consider metrics that accommodate flexible cluster shapes.   Our review and simulation study provide insight regarding several such metrics.

The Wasserstein Distance ($WD$), Hellinger Distance ($HD$), and the Extended Jensen Shannon Distance ($JSD_e$) had good accuracy in the majority of simulation scenarios, as long as cluster recovery was moderate as indicated by the Adjusted Rand Index. $JSD_e$ and $HD$ have the advantage of being bounded by $0$ and $1$, a  property that potentially allows for easier indexing of potential benchmarks for effect size. However, a caveat is that $JSD_e$ and  $HD$ flatten as they near 1 (e.g., $\geq 0.80$), which means that very small changes in value toward the upper bound may correspond to deceivingly large shifts in mean, rotation, or scale.   In contrast, $WD$ generally changed linearly, which enhances its interpretability. $WD$ also appeared to be more sensitive to differences in scale and rotation than other metrics.  For example, in Scenario 1 (mean differences between MVN clusters) the maximum $WD$ was roughly $8$, while in Scenarios 2 and 3 (differences in scale and rotation) the maximum $WD$ was roughly $4-5$ times as large.  This has important implications, as it suggests that $WD$ will generally be larger between two clusters that differ in scale or rotation, relative to two clusters that differ in mean. 

While $AB$ and $SI$ performed well in some simulation scenarios, other scenarios indicated their potential to be misleading when used for our specific questions. For example, they may may suggest cluster distances that are too large when cluster recovery is very poor, or too small even if cluster recovery is excellent. It is also important to emphasize that $AB$, $SI$, and $MD$ are not able to  distinguish differences in scale, which is a major limitation of these metrics if the interest is in quantify distances between more flexible cluster shapes.    

An important observation across both the simulation and application is that the choice of a different distance measure can lead to a different conclusion regarding which clusters have the greatest separation between them. For this reason, we recommend using multiple metrics (e.g., $WD$, $HD$, $JSD_e$) and expect that the clusters with the greatest separation will be observed across multiple metrics. This phenomenon is common in mixture modeling.  For example, the AIC, BIC, and ICL are commonly used together although they can all suggest different results.   In clustering more broadly, many different approaches can be used to suggest the number of clusters, with a wide range of potential results. 

There are limitations and caveats to our work to be considered.  First, we have not identified which levels of each metric might be considered notable in an applied context.  Linking the metrics to ballpark estimates of ``small'', ``medium'' or ``large'' effect sizes would further promote their utility~\citep{Sullivan,Cohen}.   Second, the probability-based distance measures ($HD$, $JSD_e$, $WD$) as we have presented them  are only relevant for clusters identified through mixture modeling or data simulation because they require known or estimated parameters.  We chose to focus on  mixture modeling because it facilitates identification of non-elliptical clusters, an important and emerging area of health research. However, a consequence of this focus is that the distances as proposed can not be directly applied to clusters identified through approaches that only output cluster membership (e.g., k-means algorithm).  

In future research, it may be useful to extend our work to non-model-based clustering approaches.  $WD$ may be amenable to  this adaptation because it is possible to compute it using the observed data within each cluster, rather than estimated parameters. Recent advances advancements to $WD$ may also facilitate measuring distances between clusters with different sample sizes ~\citep{sejourne}. Along these same lines, there are also existing approaches that allow for empirical estimation of $JSD$ and $HD$. We did not use empirical estimation because of the practical concern of small (and unequal) sample sizes, and the impact this might have on estimation of distance metrics.  However, future work in this area would facilitate the use of distance measures for a wider range of clustering methods.

\section*{Disclosure statement}
MLW is a consultant for Noctem, Sleep Number Bed, and Health Rhythms, unrelated to this work. 

\section*{Funding}
This work was supported by the National Institute of Mental Health under grant number R03MH116478.

\bibliographystyle{tfcse}
\bibliography{Wallace_Distance}

\end{document}